\pdfoutput=1 
\documentclass[cits]{JINST_mod}
\usepackage{textcomp} 
\usepackage{amsmath}
\usepackage{siunitx}
\usepackage{caption}
\usepackage{subcaption}
\title{Radon in the {DRIFT-II} directional dark matter TPC: emanation, detection and mitigation}

\author{
J.B.R. Battat $^a$ , J.~Brack $^b$ , E.~Daw $^c$ , A.~Dorofeev $^b$ , A.C.~Ezeribe $^c$ , J.R.~Fox $^d$ , J.-L.~Gauvreau $^d$ , M.~Gold $^e$ , L.J.~Harmon $^d$ , J.L.~Harton $^b$ , J.M.~Landers $^d$ , E.R.~Lee $^e$ , D.~Loomba $^e$ , J.A.J. Matthews $^e$ , E.H.~Miller $^e$ , A.~Monte $^d$ , A.StJ.~Murphy $^f$ , S.M.~Paling $^g$ , N. Phan $^e$ , M.~Pipe $^c$ , M.~Robinson $^c$ ,  S.W.~Sadler$^{c}$\thanks{Corresponding author.}, A.~Scarff $^c$ , D.P.~Snowden-Ifft $^d$ , N.J.C.~Spooner $^c$ , S.~Telfer $^c$ , D.~Walker $^c$ , D.~Warner $^b$ ~and L.~Yuriev $^c$ \\
\llap{ $^a$ } Physics Department, 106 Central Street, Wellesley College, MA 02481, USA\\	
\llap{ $^b$ }Department of Physics, Colorado State University, Fort Collins, CO 80523-1875 USA\\	
\llap{ $^c$ }Department of Physics and Astronomy, University of Sheffield, S3 7RH, UK\\
\llap{ $^d$ }Department of Physics, Occidental College, Los Angeles, CA 90041, USA\\
\llap{ $^e$ }Department of Physics and Astronomy, University of New Mexico, NM 87131, USA\\
\llap{ $^f$ }SUPA, School of Physics and Astronomy, University of Edinburgh, EH9 3JZ, UK\\
\llap{ $^g$ }STFC Boulby Underground Science Facility, Boulby Mine, Loftus, Saltburn-by-the-Sea, Cleveland, TS13 4UZ, UK\\
  E-mail: \email{stephen.sadler@sheffield.ac.uk}}

\abstract{Radon gas emanating from materials is of interest in environmental science and also a major concern in rare event non-accelerator particle physics experiments such as dark matter and double beta decay searches, where it is a major source of background.   Notable for dark matter experiments is the production of radon progeny recoils (RPRs), the low energy ($\sim 100$~keV) recoils of radon daughter isotopes, which can mimic the signal expected from WIMP interactions.  Presented here are results of measurements of radon emanation from detector materials in the 1m$^{3}$ DRIFT-II directional dark matter gas time projection chamber experiment.  Construction and operation of a radon emanation facility for this work is described, along with an analysis to continuously monitor DRIFT data for the presence of internal $^{222}$Rn and  $^{218}$Po. Applying this analysis to historical DRIFT data, we show how systematic substitution of detector materials for alternatives, selected by this device for low radon emanation, has resulted in a factor of $\sim 10$ reduction in internal radon rates.  Levels are found to be consistent with the sum from separate radon emanation measurements of the internal materials and also with direct measurement using an attached alpha spectrometer.  The current DRIFT detector, DRIFT-IId, is found to have sensitivity to $^{222}$Rn of 2.5~\si{\micro \becquerel \per \litre} with current analysis efficiency, potentially opening up DRIFT technology as a new tool for sensitive radon assay of materials.}

\def \alphpart{$\alpha$~particle}

\begin{document}
\section{Introduction to radon backgrounds and the DRIFT-II experiment}
\label{sec:intro}
\begin{figure}[h!tb]	
\begin{center}
\includegraphics[width=0.7\textwidth]{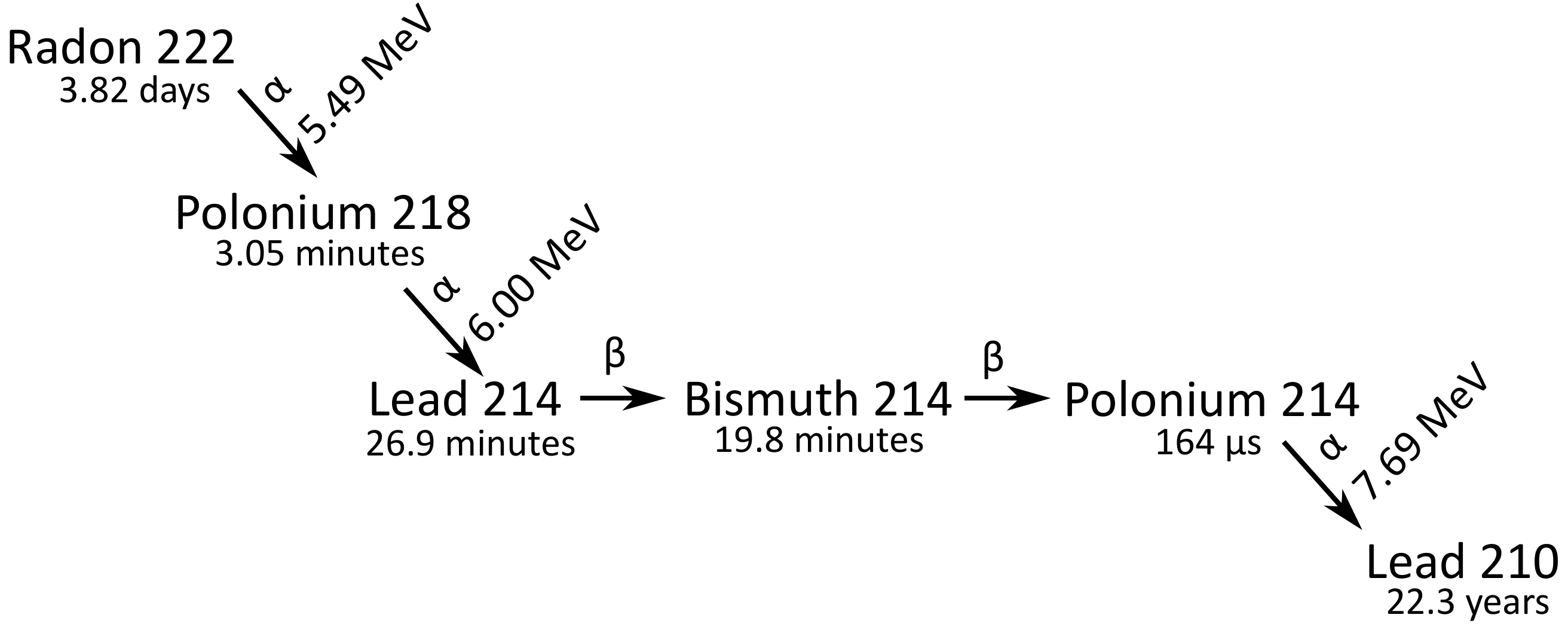}
\caption{Part of the $^{238}$U decay chain. The RAD7 radon detector (see section 2.3) uses the $\alpha$~particles from the decay of $^{218}$Po and $^{214}$Po to measure the rate of $^{222}$Rn decay in its sensitive volume.}
\label{fig:RnDecayChain}
\end{center}
\end{figure}

All materials on Earth contain some level of contamination from U chain isotopes and are a source of $^{222}$Rn gas produced by the alpha decay of $^{226}$Ra.  Together with the daughter isotopes that also decay by alpha and beta emission, this radon is the most significant source of atmospheric radiation exposure~\cite{akihiro}. We show in figure~\ref{fig:RnDecayChain} the part of the $^{238}$U decay chain relevant to this work, including half-lives of the critical isotopes and energies of the emitted alpha particles.

$^{222}$Rn is a noble gas. Its low chemical reactivity and $3.8$~day half-life allow it to migrate from the source of production inside materials, leading to its widespread distribution. Improved understanding and measurement of radon and its progeny in these processes, and of the rates of emanation from materials, is therefore of significant interest in environmental science. Radon is also a concern in fundamental physics experiments seeking to observe rare particle events, such as double beta decay searches or searches for dark matter in the form of Weakly Interacting Massive Particles (WIMPs). Radon-related backgrounds can limit sensitivity in such experiments~\cite{Beringer2012}.  Efforts to understand the radon background and to mitigate its effects have recently driven significant advances in detection technology in these fields, with potential benefits to the wider study of radon in general.  A particular example of this is development of the DRIFT (Directional Recoil Identification from Tracks)~\cite{Daw2012} series of directional dark matter experiments at the Boulby Underground Science Facility~\cite{Murphy2012}. 

For the past $16$~years the DRIFT collaboration has pioneered low-pressure gas Time Projection Chambers (TPCs) for WIMP dark matter searches. Dark matter is thought to make up $\sim 90\%$ of the mass of our Galaxy, and the determination of its nature is a major challenge in particle physics ~\cite{NRC}. The strength of WIMP interactions with baryonic matter is experimentally constrained to be extremely weak, with spin-dependent WIMP-proton interaction cross sections greater than $5.7 \times 10^{-3}$~pb ruled out for WIMPs of mass $\sim 35$~GeV/c$^2$~\cite{Felizardo2012}. Signal events are therefore expected to be extremely rare, which necessitates a very low background detector. The DRIFT experiment seeks to detect WIMPs in the galaxy by searching for the low-energy (sub-$100$~keV), nuclear recoils expected from elastic scattering in a suitable target. Unlike conventional direct dark matter detection experiments, DRIFT employs a low-pressure gas, which enables the reconstruction of the direction of WIMP-induced nuclear recoils. This provides access to the most powerful dark matter signature -- the correlation between recoil directions and the Earth's motion through the Galactic WIMP halo~\cite{Green2007, Ahlen2010}.  This approach also enables particle identification on an event-by-event basis. In particular, DRIFT allows tracking in 3D of alpha particles produced from the decay of radioactive impurities in detector materials. As described in section~\ref{sec:facility}, DRIFT can serve as a highly sensitive radon detector, with sensitivity at the \si{\micro \becquerel \per \litre} level.

\begin{figure}[htb]
\centering
\begin{subfigure}{.45\textwidth}
  \centering
  \includegraphics[width=\textwidth]{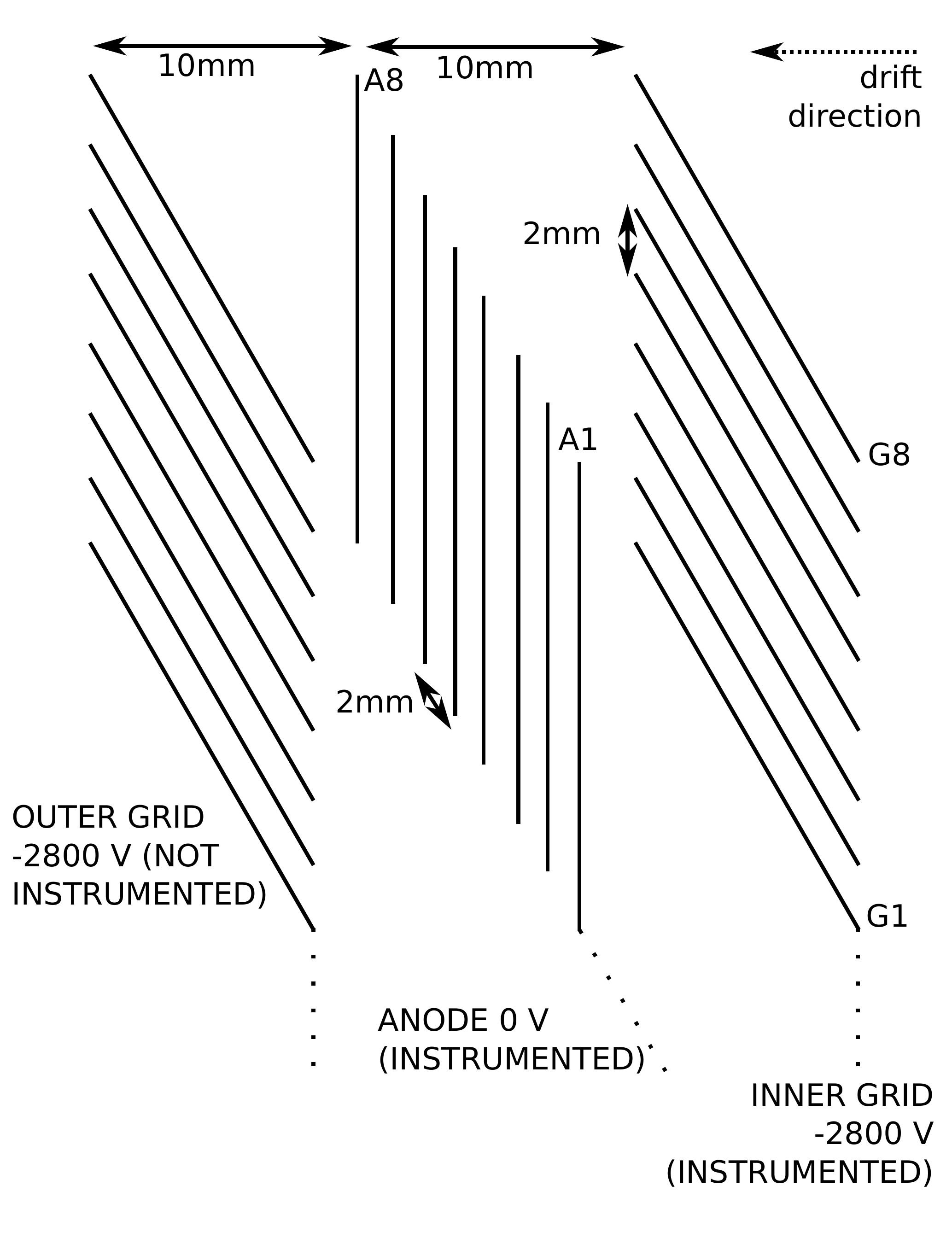}
\end{subfigure}%
\begin{subfigure}{.55\textwidth}
  \centering
  \includegraphics[width=0.8\textwidth]{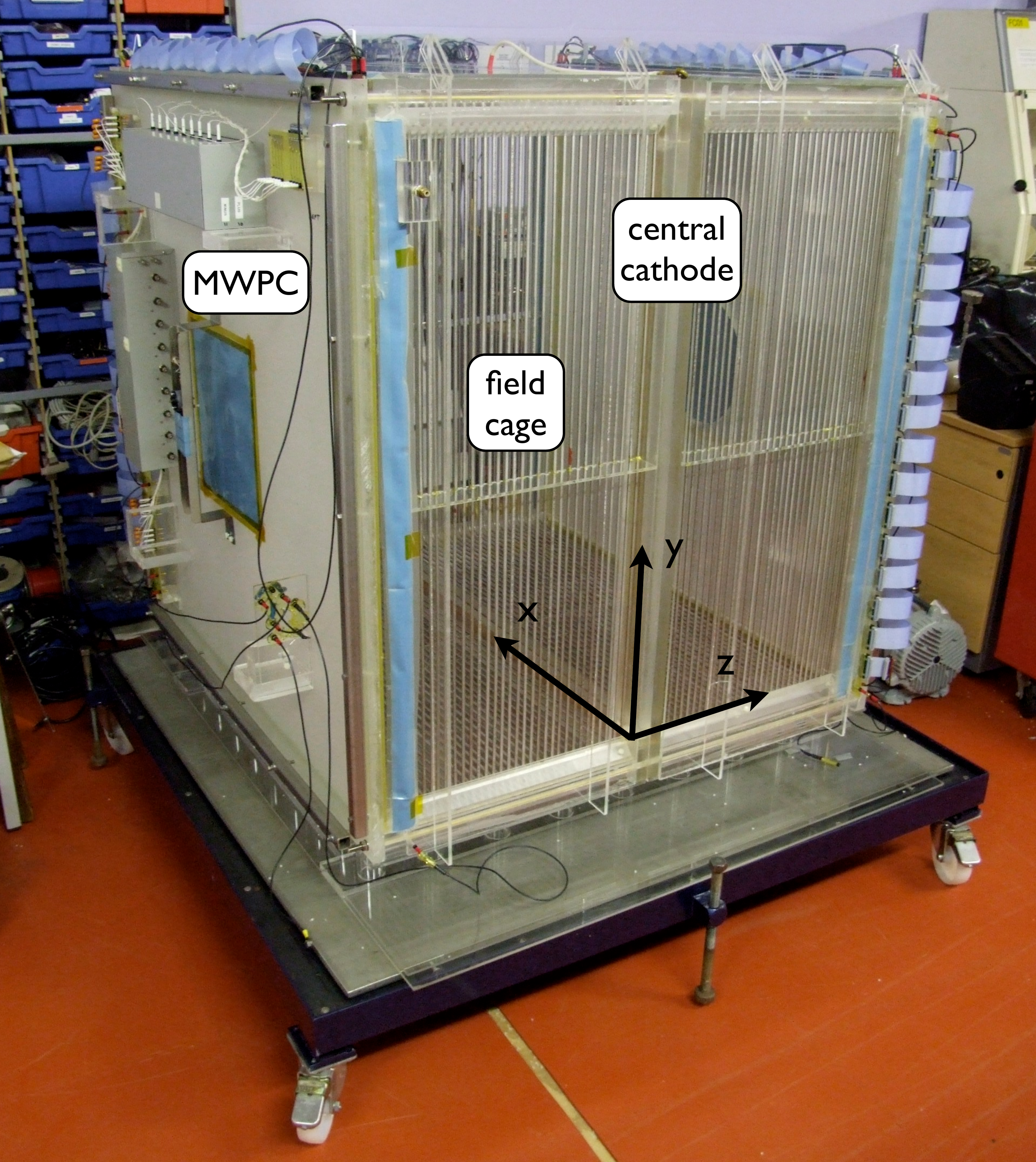}
\end{subfigure}
\caption{Left: schematic of DRIFT-IId's left MWPC detector. Right: photograph of DRIFT-IId removed from the steel vacuum vessel.}
\label{fig:drift_schematic}
\end{figure}

A photograph of the DRIFT-IId detector and a diagram of one of its MWPC detectors are shown in figure~\ref{fig:drift_schematic} as installed in the Boulby laboratory at $2805$~m.w.e. depth. DRIFT-II is a negative ion time projection chamber (NITPC) with $139$~g fiducial mass composed of 30:10~Torr CS$_2$:CF$_4$ gas.  Technical details of DRIFT-II can be found in Alner (2005)~\cite{alner2005}. Briefly, the experiment comprises a cubic steel vacuum vessel 1.5~m on a side, containing the target gas as well as two MWPC detectors with field cages of $50$~cm drift length, mounted back-to-back with a shared central cathode made from $0.9$~\si{\micro \metre} aluminised Mylar sheet. The central cathode voltage of $-30.2$~kV produces a field of $550$~\si{\volt \per \centi \metre}, which transports ionisation tracks in the fiducial volume to one of two square 1~\si{\meter \squared} MWPC detectors.  The MWPCs comprise an anode ($512$ parallel $20$~\si{\micro \metre} stainless steel wires, 2~mm pitch) sandwiched with 1~cm spacing between two perpendicular grid planes ($512$ parallel $100$~\si{\micro \metre} wires, 2~mm pitch). Ionisation electrons from the primary track are captured on the CS$_2$ with capture distance of several tenths of one \si{\milli \metre}~\cite{Ohnuki2001}, then drift as CS$_2$ negative ions, resulting in a maximum r.m.s. diffusion of  $< 0.7$~mm in both longitudinal and transverse directions, which corresponds to the thermal lower limit~\cite{Snowden2013}. Based on timing measurements of alpha particle tracks traversing the full distance between the cathode and MWPC planes, the drift speed is $59.37 \pm 0.15$~\si{\metre \per \second}.
	
Amplification with a gain of $\sim 1000$ occurs at the MWPCs by application of $2731$~V between the grid and anode planes.  To simplify the data acquisition, the anode and grid wires (measuring the x and y directions, respectively) on each MWPC are each grouped down to 8 lines by connecting every 8th wire together. In this way, 8 adjacent anode (grid) readout lines sample $16$~mm in $x$ ($y$). This is sufficient to contain all nuclear recoil tracks of interest but does mean that typical contained alpha particles, such as those from radon decay with ranges of approximately 30~cm, result in multiple signals on the readout channels (see figure~\ref{fig:gpccexample}). Counting peaks in such events allows their $x$ and $y$ ranges to be determined with $\sim 2$~mm precision. Signals are digitised at $1$~MHz so that, for the $550$~\si{\volt \per \centi \metre} drift field, one time sample corresponds to $\sim 60$~\si{\micro \metre} in the drift (z) direction. 41 (52) wires at the edge of the anode (inner grid) planes are grouped to provide an $x$-$y$ veto against backgrounds originating outside the fiducial volume. Individual events, each generating a charge track, are parameterised in terms of their energy and range using the number of ion pairs (NIPs) produced and the arrival times of these `hits'  on the grid and anode planes of the left or right (or both) MWPCs. The W-value of the 30:10~Torr CS$_2$:CF$_4$ gas mixture is measured to be $25.2\pm0.6$~eV~\cite{Pushkin2009}, and alpha tracks typically consist of $\mathcal{O}(10^5)$~NIPs.

Characterisation of the radon progeny recoil (RPR)~\cite{Burgos2007} and alpha backgrounds in DRIFT-IId over a period of $\sim 4$~years is detailed in Brack et al. (2014)~\cite{daw2014}, where progress on measuring and understanding the origin of radon backgrounds in the instrument is reported. The radon rate in the detector is determined by emanation from the internal components, and by the rate of gas flow through the vessel relative to the $^{222}$Rn half-life.  The greater the flow rate, the lower the observed rate will be. In section~\ref{sec:radonmonitoring}, we detail measurements over several years of the internal  $^{222}$Rn rate as determined from analysis of particular alpha events in the detector, using the method detailed in Burgos et al. (2008)~\cite{Burgos2008}. The reduction of this rate achieved by substituting many internal components for alternatives measured to have lower radon emanation is also presented, along with results of measurements of the radon emanation rate using an attached commercially available Durridge RAD7 alpha spectrometer. Section~\ref{sec:fulldetrn} presents measurements of the individual radon emanation rates from detector materials relevant to the changes made to the experiment. The rates are found to be compatible with those determined from the analysis of DRIFT data described in section~\ref{sec:radonmonitoring}. Section~\ref{sec:facility} presents measurements of DRIFT-IId's sensitivity to $^{222}$Rn at the \si{\micro \becquerel \per \litre} level.
 
\section{Radon Monitoring in DRIFT}\label{sec:radonmonitoring}
The design of DRIFT-II, with thin-film central cathode and two back-to-back MWPC detectors (see section~\ref{sec:intro}), allows identification of several classes of radon-related alpha background as detailed in Burgos et al. (2008)~\cite{Burgos2008}.  Notable here are so-called `gold-plated cathode-crossers'  (GPCCs) that, though not a background for dark matter searches, can provide an unambiguous tracer of radon in the experiment.  The main characteristic of GPCCs is that charge is detected in time coincidence on both sides of the detector, and that both tracks start and end within the fiducial volume. GPCCs are thus `fully contained'  events that originated in the bulk of the gas and crossed the central cathode plane. The low ($\sim 1$~Hz) raw event rate and analysis cuts make time-coincident events from separate sources extremely unlikely ($< 1$ coincidence every five years).

\begin{figure}[h!tb]
\begin{center}
\includegraphics[width=0.6\textwidth]{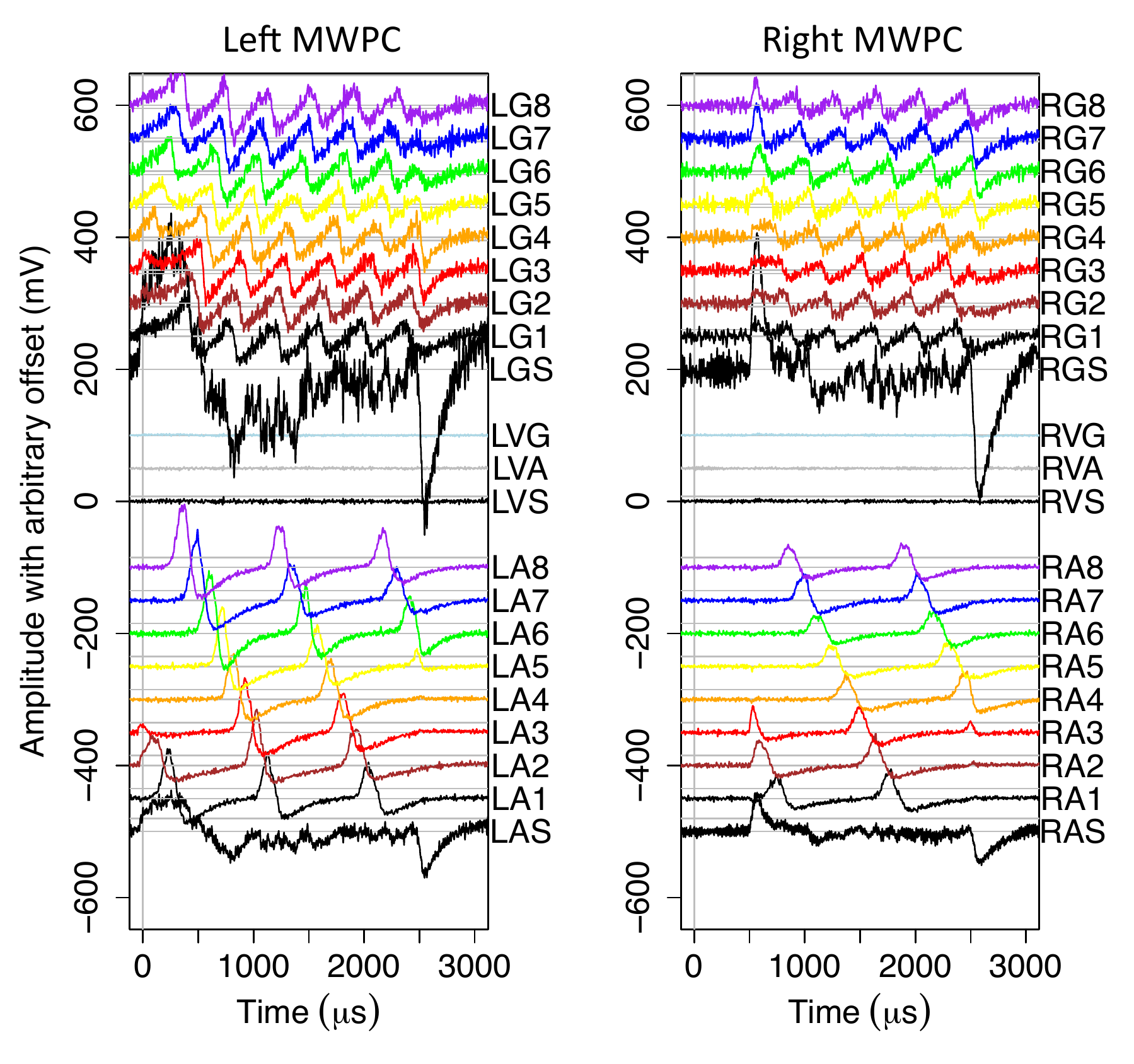}
\caption{A candidate GPCC event in DRIFT-IId showing the characteristic repeated pattern of hits on all lines along with the late-time-coincidence of the ends of the tracks (the latest hits) on either side at $\sim 2400$~\si{\micro \second}. In the waveform labels to the right of each plot, "L" and "R" refer to the left and right MWPC, "G" and "A" refer to a grid and anode line, "V" refers to a veto line, and "S" means the sum. For example, "LGS" means the sum of all grid lines in the left MWPC.}
\label{fig:gpccexample}
\end{center}
\end{figure}

Figure~\ref{fig:gpccexample} shows an example GPCC with output from the two MWPCs, divided between 8 grid outputs (top lines) and 8 anode outputs (bottom lines) plus veto signal (centre three lines). 
Further details of such plots and the analysis process used can be found in Daw et al. (2012)~\cite{Daw2012}.  In this example, charge is clearly deposited in both left and right MWPCs. The grouping together of every eighth MWPC wire results in periodic charge deposition on each line, where multiple hits on a single channel correspond to the alpha particle traversing multiple $16$~mm `octaves' of the MWPC readout.  To qualify as a GPCC the particle must also be fully contained, and hence not cross the veto region.  Indeed, no veto signal (event on the centre three lines in figure~\ref{fig:gpccexample}) is seen. The only particles capable of depositing a detectable amount of charge in a fully contained track crossing both sides of the detector are alpha particles produced in the decay of an unstable atom in the gas, within the fiducial volume. Furthermore, the parent particle must be neutral, since positively (negatively) charged species drift to the cathode (anode) on $\sim 10$~\si{\milli \second} timescales.

A possible exception exists in the form of alpha particles entering the fiducial volume through either of the MWPCs and thus evading the vetoes, however in the energy range of interest, such particles do not have sufficient range to reach the central cathode. Based on this, and knowing the detector's efficiency for detecting GPCCs from a given radioactive species (see section~\ref{sec:2.2}), the rate of decays of that species inside the vacuum vessel can be inferred. Combining this inferred decay rate with knowledge of the flow rate of gas through the vessel and the species'  half-life, the rate of appearance of that species inside the vessel under equilibrium conditions can be calculated. 

\subsection{Analysis of GPCC events}
\label{sec:2.2}
To illustrate the analysis process for GPCCs, we consider here a typical dataset of 6.6~live-days from April 2012, taken immediately preceding a radon emanation measurement made using an attached alpha spectrometer described in section~\ref{sec:rad7}.  Data were collected immediately following a full evacuation of the vessel using an Edwards XDS10 scroll pump (reaching an ultimate pressure of $\sim 0.2$~Torr), and subsequent re-starting of the gas mixture's flow. A basic set of cuts were applied to ensure that 1) the waveforms were within the range of the digitisers; 2) there was no activity outside the region of interest (ROI) between -500 and 5000~\si{\micro \metre}; 3) there was no veto activity. The cuts given in Table~\ref{tbl:gpcccuts} were then applied to the data in order to select GPCC candidates from the remaining events. A total pre-cuts trigger rate of $1.22$~Hz was observed, with $689$ GPCCs surviving the cuts for an average rate of $104 \pm 4$ events per day. These events may originate from any uncharged radioactive species generating alpha particles in the fiducial volume.

\begin{figure}[bt]
\begin{center}
\includegraphics[width=0.45\textwidth]{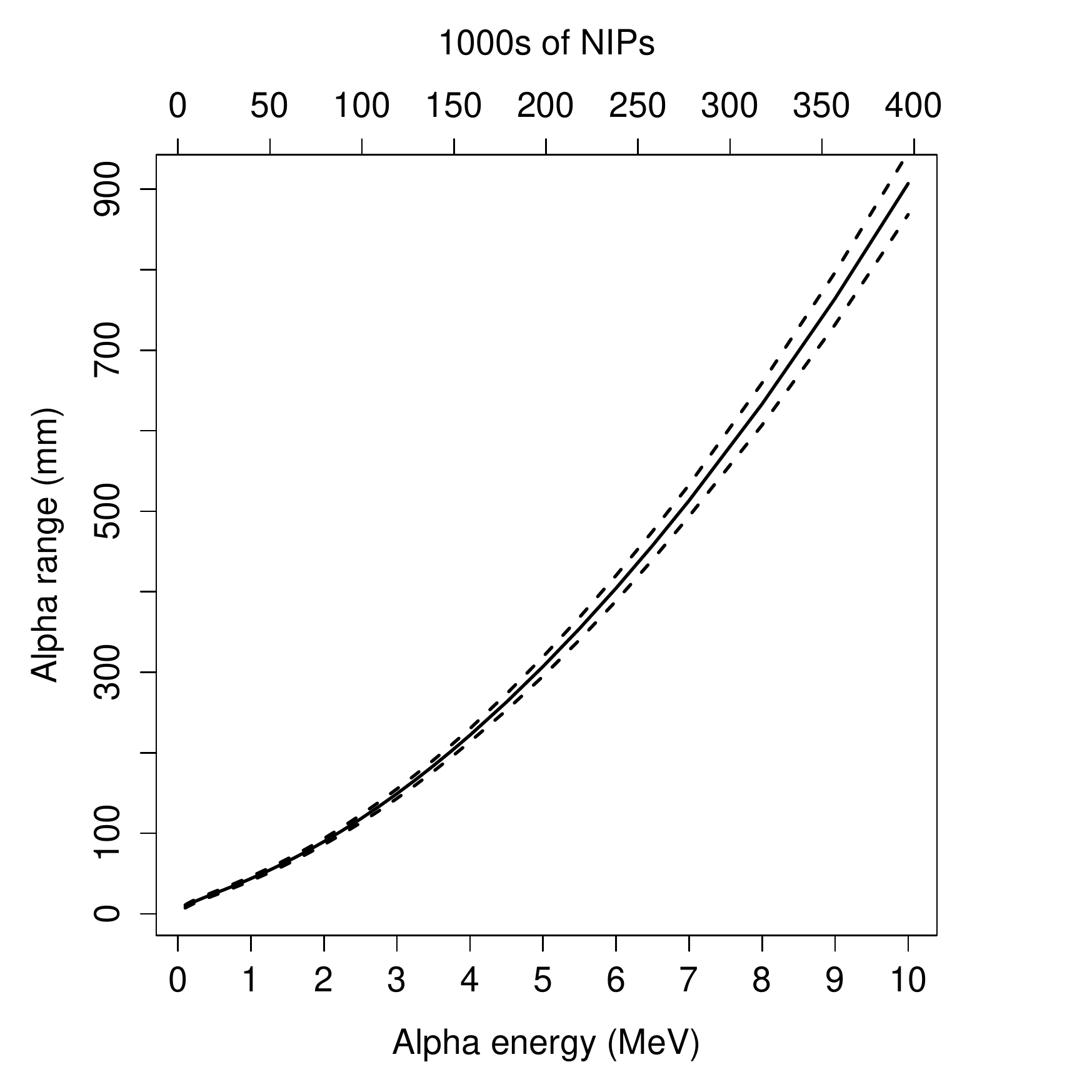}
\caption{Alpha particle range in 30:10 Torr of CS$_2$:CF$_4$ vs. initial alpha particle energy, calculated using the Monte Carlo simulation program SRIM 2011~\cite{Ziegler1985}. Dashed lines represent longitudinal straggling (see text).}
\label{fig:SRIMrange}
\end{center}
\end{figure}

To unambiguously identify a given alpha track as originating from $^{222}$Rn decay, it is important to separate out the different alpha species. This is achievable in DRIFT by reconstructing the tracks in 3D and hence calculating the 3D range of the GPCC alpha particles~\cite{Burgos2008}. Radioactive species emit alpha particles of characteristic energies, with the higher energy particles travelling further before `ranging out'. Using the Stopping Range of Ions in Matter program (SRIM 2011)~\cite{Ziegler1985}, a table of alpha energy vs. range in the gas was computed, and linear interpolation used to extract the range and straggling of alpha particles of energies between $0.1$ and $10$~MeV in 30:10~Torr CS$_2$:CF$_4$ gas (see figure ~\ref{fig:SRIMrange}). Straggling is defined as the standard deviation of the simulated particle ranges in the direction parallel (longitudinal straggling) and perpendicular (lateral straggling) to the initial simulated alpha particle trajectory. As shown in figure~\ref{fig:RnDecayChain}, the alpha from $^{222}$Rn decay carries $5.49$~MeV corresponding to $354\pm14$~mm in DRIFT. For reference, other decays of interest are: $^{220}$Rn decay at $5.29$~MeV ($335 \pm 13$~mm), $^{218}$Po at $6.00$~MeV ($404\pm16$~mm), and $^{216}$Po with $6.78$~MeV ($489 \pm  19$~mm). In practise all measured GPCC ranges were slightly shorter than those predicted by SRIM, primarily due to energy loss in the thin-film cathode.
\begin{table}
\begin{centering}
\resizebox{0.75\textwidth}{!} {
\begin{tabular}{| c | c | c | c |} \hline
Name 					& 	Description 												&	Acceptance (\%)	\\ \hline
Basic cuts					&	Ensure events are analysable.									&	62				\\ 
Charge on both sides		&	Requires there to be channels hit on each side.					&	5 				\\ 
Eight channels hit			&	Ensures $\Delta x > 16$~mm on both sides.		 				&	24				\\ 
Delta z cut				&	$\Delta z  >  50$~mm. Removes remaining sparks					&	98				\\ 
						&	and ensures that the track can be well-measured.					&					\\ \hline
\end{tabular}
}
\caption[Cuts used to select GPCC events.]{Cuts used to select GPCC events. Acceptances are calculated based only on the events that make it through the preceding cuts, and so depend upon the order in which they are applied. `Sparks' are impulse charge depositions. \label{tbl:gpcc_cuts}}
\label{tbl:gpcccuts}
\end{centering}
\end{table}

The event-by-event calculation of GPCC track ranges requires separate analyses for the $x$, $y$ and $z$ dimensions.  The $x$ extent is reconstructed by counting wire hits on the anode channels. 
The method for counting hits uses the time derivative of the anode waveforms, after smoothing with a Savitzky-Golay smoothing filter to remove noise peaks which may otherwise be counted as true charge depositions (the `SD' waveform)~\cite{Savitzky1964}. The time derivative is defined as the difference in amplitude between waveform samples separated by 5~\si{\micro \second}. Scaling the smoothing width in proportion to the time-extent of the event ensures that valid fast events are not erroneously `smoothed out'. Each time a channel's SD waveform drops below a fixed threshold, 2~mm is added to the track's $x$~range. The $y$ range was calculated by applying a similar algorithm to the grid channels. Finally, the $z$ range ($\Delta z$) of GPCCs was calculated from:
\begin{equation}
\label{eqn:deltaz}
\Delta z = (\Delta t_{left} + \Delta t_{right} ) v_{drift} \text{.}
\end{equation}
Here $v_{drift}$ is the CS$_2$ negative ion drift velocity ($59,445$~\si{\milli \metre \per \second} for this run), and $\Delta t_{left}$ and $\Delta t_{right}$ are the duration of the charge deposition on the left and right MWPCs, respectively.

\begin{figure}[h!tb]
\begin{center}
\includegraphics[width=0.55\textwidth]{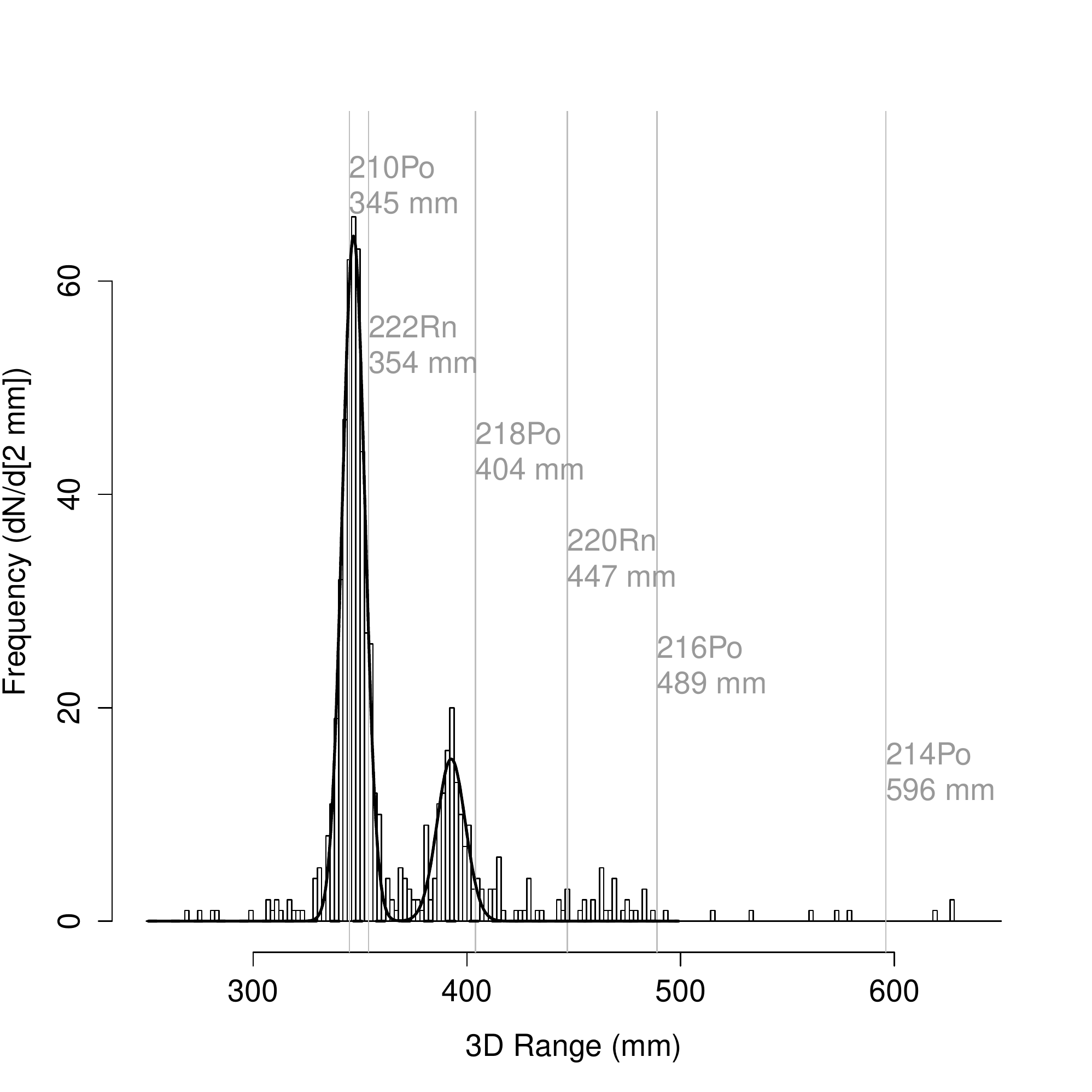} 
\caption{Reconstructed 3D ranges for GPCC events during $6.6$~live-days. \alphpart s from $^{222}$Rn decay populate the peak at $\sim 347$~mm, whilst those from $^{218}$Po decay appear around $393$~mm. Ranges from SRIM of other isotopes of interest are overlaid in grey. All are systematically higher than the peaks in data due to energy loss in the central cathode.}
\label{fig:gpccrange}
\end{center}
\end{figure}

Based on this range analysis, we show in figure~\ref{fig:gpccrange} a histogram of the 3D range of GPCCs in the example dataset, fit with a double Gaussian profile. The peaks at $347$ and $393$~mm are $ \sim 5$\% lower than the alpha ranges for $^{222}$Rn and $^{218}$Po decay, respectively, shown in Figure~\ref{fig:SRIMrange}. This small discrepancy is due to energy loss as the particles pass through the central cathode, which is discussed later. There is no evidence for any of the $^{220}$Rn (thoron) peaks in the distribution. The areas under the Gaussian peaks divided by the detector livetime give the average rates of detected GPCC events for the respective species.

\subsection{Efficiency corrections and the effect of gas flow}
\label{subsec:alpharange}
The analysis above yields the measured contained GPCC alpha rate due to uncharged $^{222}$Rn and $^{218}$Po. The calculation of an `efficiency factor' is required to convert the measured GPCC rate to the decay rate in the whole vessel for each species, for which a Monte Carlo simulation code was written. For each species, a mono-energetic population of alpha particles was generated with a uniformly distributed random position in a  $1.5 \times 1.5 \times 1.5$~m volume representing the DRIFT-IId vacuum vessel. The particles were then transported with range as calculated by SRIM 2011 (see Figure~\ref{fig:SRIMrange}) and with orientations drawn randomly from uniform distributions in $\cos \theta$ and $\phi$, where $\theta$ is the polar angle measured up from the $z$-axis toward the central cathode, and $\phi$ is the azimuthal angle around the $z$ axis, measured from the $x$ axis. Energy loss in the cathode was modelled using SRIM as a function of $\theta$ and the alpha particle's energy at the point of entry into the cathode. GPCC selection cuts requiring a) full containment, and b) one cathode crossing, were applied to the Monte Carlo output. The ratio of the number of events satisfying these criteria to the total number of events generated ($1 \times 10^6$) yields the `geometric efficiency factor', which was found to be $0.0274 \pm 0.0017$ for $^{222}$Rn, and $0.0298 \pm 0.0018$ for $^{218}$Po, the difference coming from the differing ranges of the two decay alpha particles.

\begin{figure}[h!tb]
\begin{center}
\includegraphics[width=0.9\textwidth]{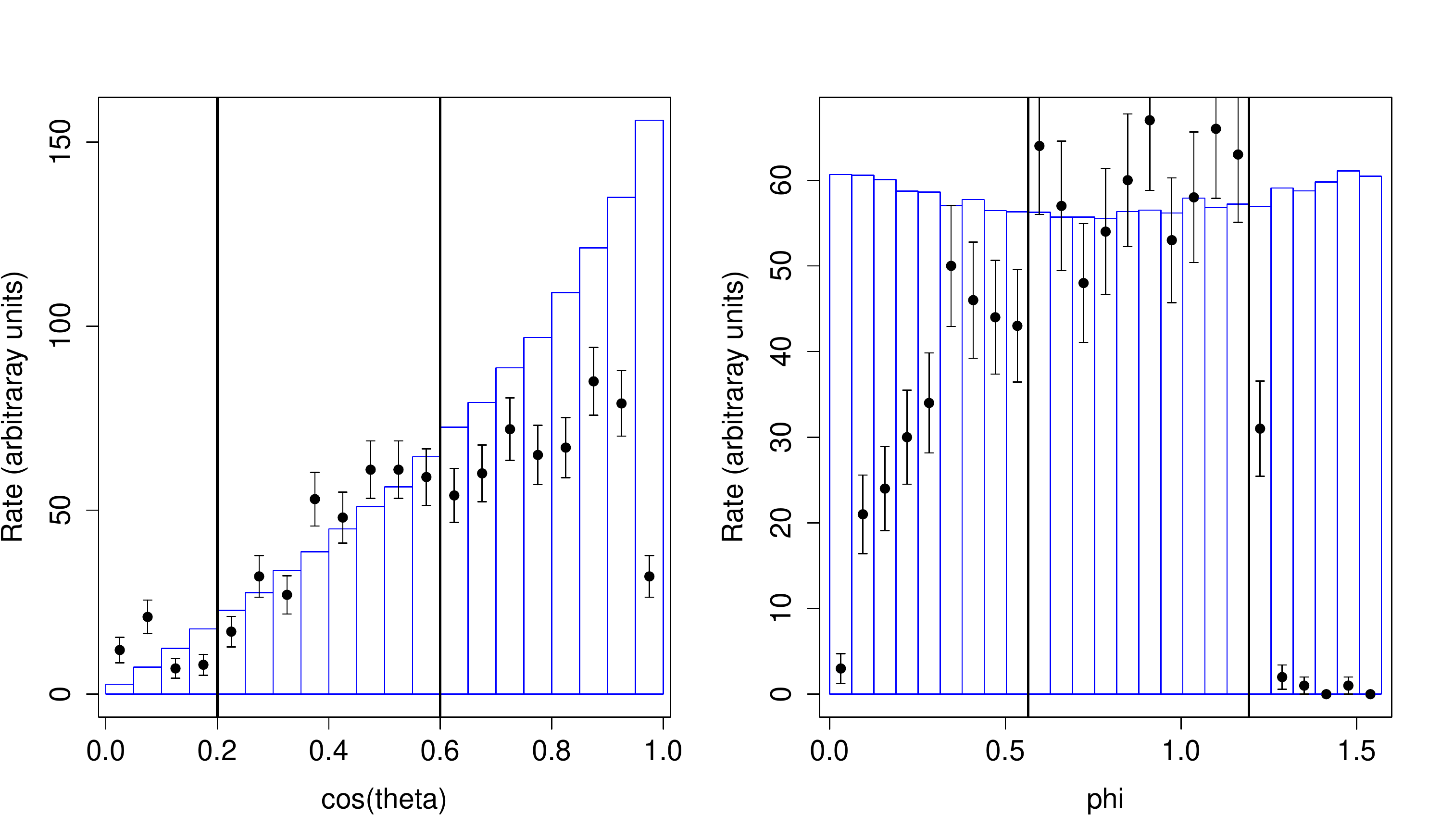} 
\caption{$\cos \theta$ and $\phi$ distributions for data (black points) and Monte Carlo (blue histogram) events passing the GPCC selection cuts. The Monte Carlo distribution was normalised to match the data between the black vertical lines.}
\label{fig:thetaphi}
\end{center}
\end{figure}

The $\cos \theta$ and $\phi$ distributions of events fulfilling the above criteria were normalised to the data in a restricted range at moderate angles, where the efficiencies were assumed to be $100 \%$ (see figure~\ref{fig:thetaphi}). The ratio of the areas under the data and simulation curves was used to calculate angular efficiency factors in $\theta$ and $\phi$, which were $<1$ due to DRIFT's difficulty reconstructing events perpendicular to any of the three axes. The combined angular efficiency factor was found to be $0.474 \pm 0.080$ which, when multiplied by the geometric efficiency factors, yields an overall efficiency factor of $(1.30 \pm 0.23) \%$ for $^{222}$Rn and $(1.41 \pm 0.25) \%$ for $^{218}$Po. This is the fraction of decays occurring in the whole vacuum vessel that would be successfully identified as such by the DRIFT detector and retained by the analysis.
	
Under equilibrium conditions with no gas flow and the system isolated for a time $t \gg t_{\frac{1}{2}}$ (the half-life of the species in question), the calculated event rates equal the appearance rates of the respective species in the vessel. However, normally the gas is flowed at a constant rate of one vessel volume change per day ($590$~g/day), which causes  a lowering of the measured alpha rate by an amount dependent upon the species half-life. This can be accounted for using the method of Burgos et al. (2008)~\cite{Burgos2008}:
\begin{equation}
\label{eqn:simpleDvsE}
D_x = A_x  \frac{1}{1+\frac{\tau_e}{\rho}} \text{,}
\end{equation}
which is valid for equilibrium conditions with constant flow. For a given species $x$, $D_x$ is the measured rate of decay in the vessel, $A_x$ is the appearance rate, $\tau_e$ is the decay time $\left( \frac{t_{1/2}}{\ln(2)} \right)$, and $\rho$ is the characteristic `flush time' of the vessel. $A_x$ represents the sum of processes that are adding species to the vessel volume. In the case of $^{222}$Rn, this is emanation from radium-containing detector materials, whereas the $^{218}$Po is produced directly in the volume from the decay of $^{222}$Rn itself.  The flush parameter $\rho = 1$~day when flowing at the nominal rate. 

Together, equation~\ref{eqn:simpleDvsE} and the Monte Carlo-derived efficiency factors allow calculation of $A_{x}$ for the detected species in the vessel, and hence the appearance or emanation rates in atoms/s.  Typically $A_{x}$ is calculated by discarding the first few days of data, which is `out of equilibrium', and plotting a histogram of the range of alpha particles in remaining data. The range peaks, for instance $^{222}$Rn and $^{218}$Po shown in figure~\ref{fig:gpccrange}, are then fitted with Gaussians to extract event rates for  $^{222}$Rn and $^{218}$Po, including a livetime correction.  For the example data this yields $64.2\pm2.8$ and $19.1\pm3.1$~events per day, respectively. Dividing by the relevant species' efficiency factor yields the total rate of decay of that species in the entire vessel: $D_{Rn} = 3.43 \pm 0.63$~decays/min and $D_{Po} = 0.93 \pm 0.25$~decays/min. Finally, equation~\ref{eqn:simpleDvsE} allows the equilibrium appearance rates of $^{222}$Rn and $^{218}$Po to be calculated, yielding in this case  $^{222}$Rn: $22.33 \pm 4.06$~atoms/min, $^{218}$Po: $0.93 \pm 0.25$~atoms/min.

Only uncharged radioisotopes can contribute to the measured GPCC rates, since charged particles immediately drift to either the central cathode or MWPC grid wires, and hence are unable to produce a GPCC event. Whereas the parent population of $^{222}$Rn is produced $100\%$ uncharged, the daughter $^{218}$Po has a neutral fraction that is a strong function of several difficult-to-control variables such as relative humidity and trace aerosol concentration, as well as absolute pressure. These have been investigated by, for example, Hopke~\cite{Hopke1996}. The ratio of the equilibrium rate of decay of $^{222}$Rn to the equilibrium rate of appearance of $^{218}$Po provides an estimate of the neutral fraction of $^{218}$Po, which was found to be $(27 \pm 9) \%$, the large uncertainty being a consequence of the low rate of $^{218}$Po GPCCs. This is in agreement with the value of $22\%$ found in Burgos et~al.~(2008)~\cite{Burgos2008}.
	
\subsection{Independent confirmation of radon emanation rate}
\label{sec:rad7}
The process above provides a powerful tool for determining the emanation rate of radon species into the vessel by direct analysis of DRIFT data and is key to the results of  section~\ref{sec:fulldetrn}.  In order to build confidence in the technique, an independent direct measurement of radon emanation inside the vessel was made using a commercially available Durridge RAD7 radon detector~\cite{Grodzins2009}.   Following a DRIFT run, the vessel was pumped down for one week to extract radon and any CS$_2$ and CF$_4$ outgassed from the detector materials (notably the acrylic frame), reaching the ultimate pressure of the system ($\sim 0.2$~Torr). The vessel was then sealed and left for a further week to allow the detector to emanate radon from the decay of $^{226}$Ra near the surface of detector materials, as it would during normal running. The vessel was backfilled with 8 Torr of dry N$_2$ then pumped, along with the emanated radon and residual gasses, into a smaller ($35$~L) vessel. The backfill pressure was chosen such that after transfer the pressure in the smaller vessel was 1~bar, the nominal operating pressure of the RAD7. Concentrating the gas in this way improves sensitivity by the ratio of the two volumes, here a factor of $\sim100$ compared with sampling directly from the DRIFT vessel.

\begin{figure}[h!tb]
\begin{center}
\includegraphics[width=0.6\textwidth]{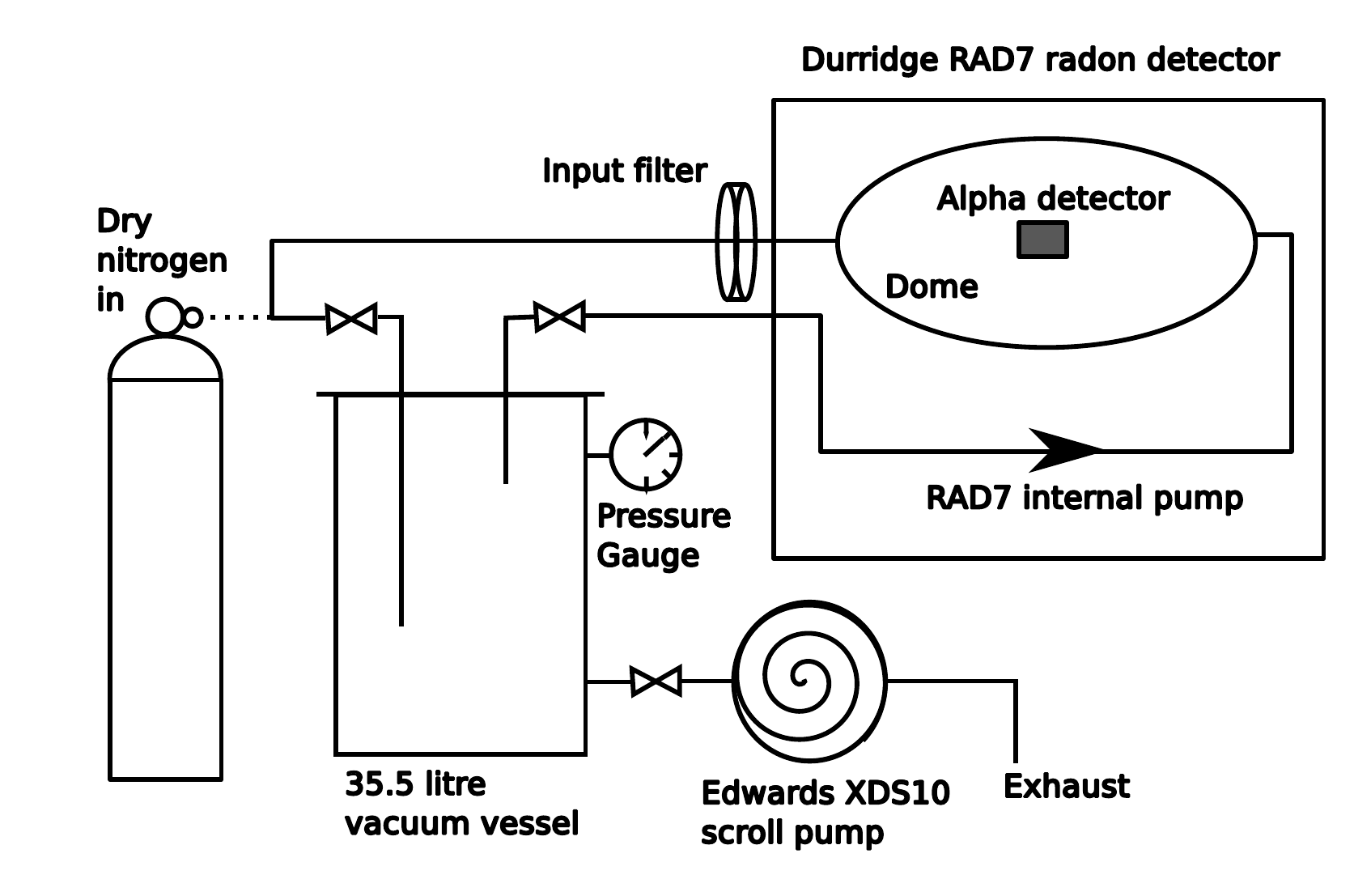}
\includegraphics[width=0.32\textwidth]{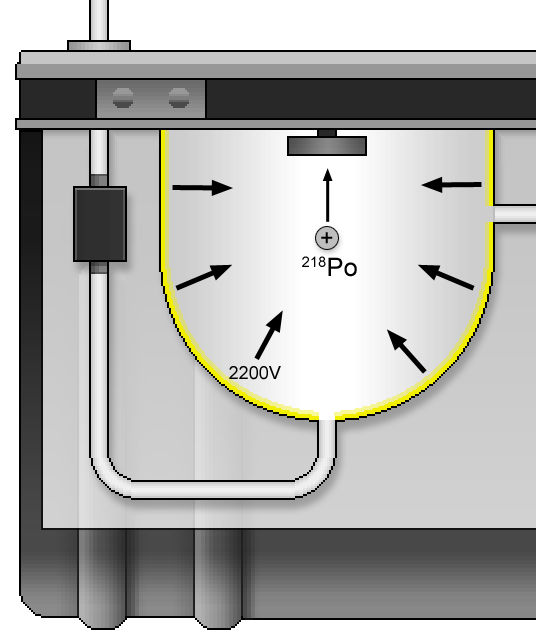}
\caption{Left: sketch of the radon emanation apparatus. Right: detail of the RAD7 sensitive volume (image credit: Durridge).}
\label{fig:RnRig}
\end{center}
\end{figure}

The RAD7 was connected to the apparatus as shown in figure~\ref{fig:RnRig}. Each run consisted of 12 measurements of the radon concentration over four hours (`cycles'), during which the RAD7's internal pump circulated the gas through its $1.3$~\si{\litre} alpha spectrometer detector at a measured rate of $1.075\pm0.012$~m$^3$s$^{-1}$. A \si{\micro \metre} mesh filter at the input ensured no charged or reactive radioisotopes were admitted, which could otherwise contribute background counts to the measurement. Charged daughter particles produced in radon decays in the sensitive volume were swept to the silicon diode radiation detector at the centre of the dome under the influence of the electric field between the $0$~V silicon detector and the $2200$~V  dome. Due to the geometry of the detector, there was then a $50\%$ chance that the radon daughter's subsequent alpha decay would deposit the alpha's energy in the silicon detector, and a $50\%$ chance that the alpha would be emitted into the gas. The silicon detector is large enough to fully contain those alpha particles which enter it, thus guaranteeing that their full energy was measured~\cite{Grodzins2009}. 

The RAD7 measures the energy of plated-out $\alpha$~particles with a precision of $50$~keV in the energy range 0 -- 10~MeV. The number of counts in two energy windows is summed: 5.40 -- 6.40~MeV to catch the $6.0$~MeV \alphpart s from $^{218}$Po , and 7.2 -- 8.2~MeV for the $7.69$~MeV $\alpha$ decay of $^{214}$Po (see Figure~\ref{fig:RnDecayChain}). The result is corrected for $^{212}$Po contamination, then divided by the detector livetime, and finally by two to give the equilibrium count rate due to radon decay of a given daughter species.  Finally, a manufacturer-determined calibration constant is used to calculate the total equilibrium radon concentration in \si{\becquerel \per \meter \cubed}, which forms the output for each of the RAD7's 12 cycles.

The 12 measurements of radon concentration in the sensitive volume were subject to several corrections, as follows. Firstly, the intrinsic background of the RAD7 was subtracted, which was measured by the manufacturer to be $0.2$~\si{\becquerel \per \meter \cubed}. The RAD7 calculates the specific activity inside its sensitive volume, A$_{s}$, in Bq/m$^3$ from the measured rate of counts in the $^{218}$Po and $^{214}$Po energy windows via a factory-determined calibration constant ($\sim 0.0135$~\si{cpm \per  (\becquerel \per \meter \cubed)}), and the total activity `A' in the system shown in figure~\ref{fig:RnRig} is thus A = A$_{s}$ $\times$ (V$_{chamber}$ + V$_{RAD}$) where V$_{chamber}$ and V$_{RAD}$ in this case are the volumes of the 35~L chamber and RAD7, respectively.

A correction for the effect of the relative humidity ($RH$) at the time of the measurement was applied to tests in which $RH > 15\%$ . Being a polar molecule, water vapour attracts ions and hence reduces the instrument's sensitivity by preventing radon daughters from reaching the detector. The RAD7 monitors the $RH$ of the sample gas, allowing its effects to be accounted for. This was done using equation~\ref{eqn:rh}, which is the room-temperature approximation of the correction empirically determined by Durridge.
\begin{equation}
\label{eqn:rh}
{A}_{\text{after}} = A_{\text{before}} \times \frac{100}{116.67-1.1\times RH} \text{,}
\end{equation}
where $A_{\text{before}}$ and $A_{\text{after}}$ represent Rn activity before and after the RH correction, respectively.

Next, an emanation time adjustment is needed. During the emanation period the rate of radon emanation into the volume remains constant, whilst the rate of radon decays increases in proportion to the number of radon atoms in the volume, approaching equilibrium where the rate of decay is equal to the rate of emanation. Since it is not possible to wait until  t$=\infty$, the following equation was used to correct the measured activity to that which would have been measured under equilibrium conditions, which is itself equal to the emanation rate under these conditions:
\begin{equation}
\label{eqn:emanationcorrection}
R_{\text{em}} = A_{\text{meas}} \times \frac{1}{1- \exp \left(\frac{-t}{t_{e}}\right)} ,
\end{equation}
where $t$ is the emanation time and $t_e = 5.52$~days is the time for radon to decay by a factor of 1/e~\cite{Colle1995}.

Finally, an emanation time-dependent background arises from radon emanating from the apparatus shown in Figure~\ref{fig:RnRig}, such as the o-ring seals on KF vacuum fittings. Flushing for 15~minutes with dry N$_{2}$ immediately before each test ensured that no previous radon contributed to the background.  However, new radon will emanate from these components to mix with the `signal' radon. This was corrected for by making a background measurement with only dry N$_2$ inside the emanation vessel, following exactly the same procedure as for a normal run, and the result subtracted from the emanation rate. The measurement sensitivity was defined as the radon emanation rate required to produce 3 counts in a $4$~hour cycle. Assuming $RH < 15\%$, $t = 7$~days, and perfect background subtraction, this was calculated to be $1.2$~atoms/min.

Using the technique above, measurements can be made of the intrinsic radon levels in the DRIFT-II vessel independent of the GPCC analysis of section~\ref{sec:2.2}. For instance, for the example data here, subtracting the zero-emanation control yields a value for the radon emanation rate into the vacuum vessel of $15.42 \pm 1.32$~atoms/min, which is lower than the value of $22.33 \pm 4.06$~atoms/min calculated from the measured GPCC rate at this time (see section~\ref{subsec:alpharange}). Possible explanations for this are outgassed CS$_2$ molecules in the RAD7, which would act to reduce the efficiency of the measurement by the same mechanism as the water vapour discussed earlier, and incomplete pumping of the radon from the DRIFT-II vessel to the $35$~L vessel. Nevertheless, the fact that two completely independent measurement techniques produced comparable radon emanation rates lends weight to both measurements.

\subsection{Radon rate vs. time}
\label{subsec:radonrate}

\begin{figure}[h!tb]
\begin{center}
\includegraphics[width=0.56\textwidth]{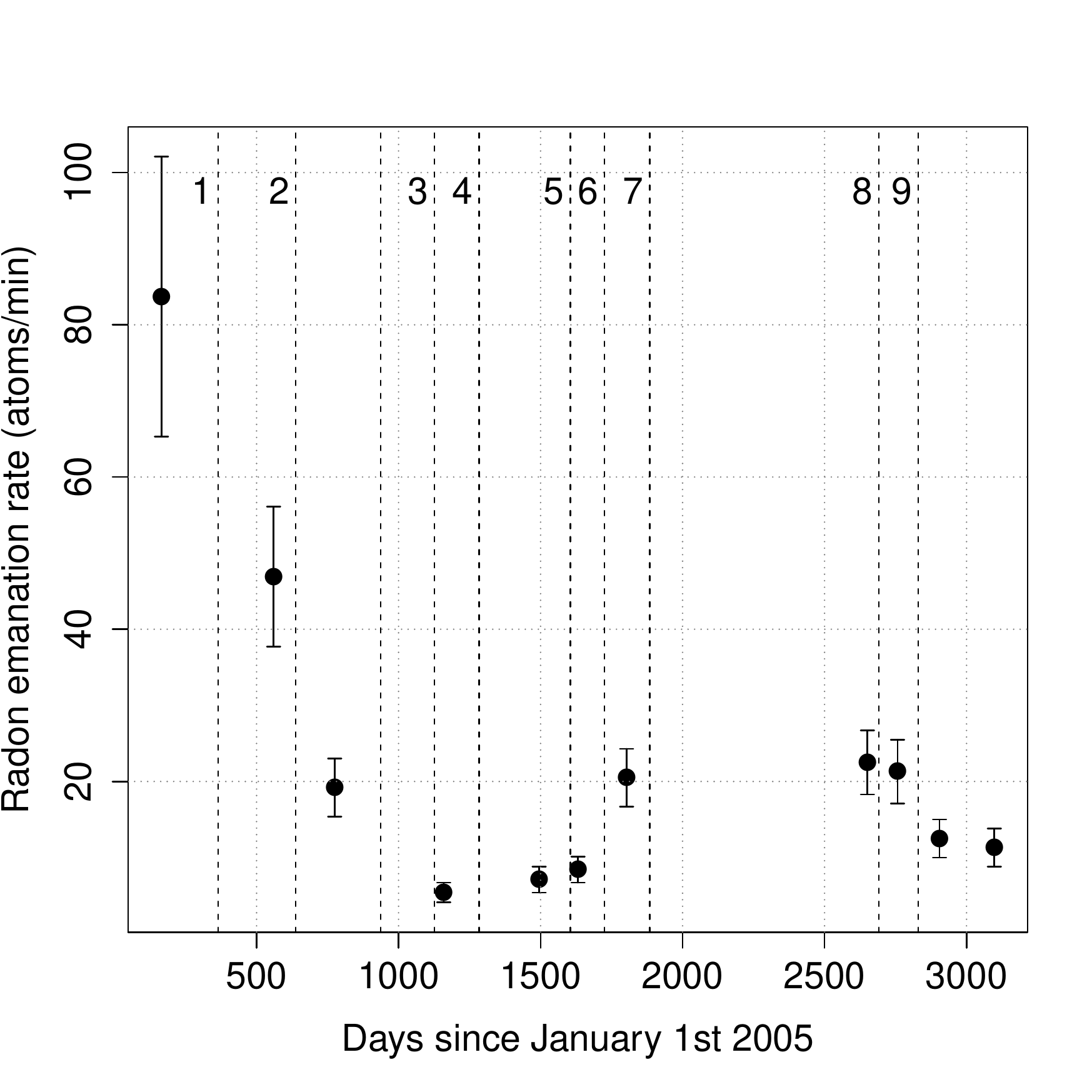}
\caption{Radon emanation rate vs. time. Dashed vertical lines represent changes to the detector configuration (see text for details).}
\label{fig:radonrate}
\end{center}
\end{figure}

The work above provides confidence that the GPCC analysis of DRIFT data can be used to monitor the intrinsic radon emanation rate into the vessel at any time. This will be important in section~\ref{sec:fulldetrn}, where we show how this measurement can be linked to the radon emanation rates from individual component materials in the detector, and how this changes as these materials have been substituted over time by materials with lower emanation rates. As a precursor to this, figure~\ref{fig:radonrate} shows the GPCC-inferred radon emanation rate in DRIFT-II over the past $9$~years. Vertical dashed lines represent changes to the detector configuration that were expected to affect the total radon emanation rate (see Brack et al. (2014)~\cite{daw2014} for more details). From left to right, these are: 1) radon refit (cables stripped and inner detector sealed); 2) signal cables replaced with Teflon-insulated equivalents; 3) central cathode etched in nitric acid; 4) MWPCs etched in the same way; 5) gas target changed from $40$~Torr CS$_2$ to 30:10~Torr CS$_2$:CF$_4$; 6) right MWPC swapped for one that had not been etched; 7) right MWPC swapped back and thin-film aluminised Mylar cathode installed; 8) radio-pure thin-film cathode installed; 9) silicone bungs and Teflon-insulated HV cables installed. Overall, a reduction in the radon emanation rate by approximately a factor of $10$ has been achieved.

\section{Radon Emanation Results from DRIFT Internal Materials}
\label{sec:fulldetrn}

From knowledge of the total rate of radon emanation into the DRIFT vacuum vessel as derived in section~\ref{sec:2.2}, two questions arise: which detector components contribute significantly to this rate, and are there radon-cold alternatives? To address these questions, a separate simple radon emanation facility was developed and used to assay all internal components.

\subsection{Radon emanation facility and analysis process}

The apparatus shown in figure~\ref{fig:RnRig}, based on a commercial RAD7 radon detector, was used once again, but this time connected in a closed loop to one of three (one $3.5$~\si{\litre}, and two $35.5$~\si{\litre}) purpose-built radon emanation chambers into which samples of materials can be placed. All components for this apparatus were themselves chosen for likely low radon emanation. A sample of material to be tested was placed into one of the three emanation chambers, then pumped down to $<0.1$~Torr to remove ambient radon, neutral daughters and any other gases, such as water vapour, which might reduce the detection efficiency. The smallest vessel that fit the sample was chosen, in order to maximise the radon concentration. The chamber and sample were left for 48~hours to allow outgassing, then evacuated and sealed once more, the start time of the radon emanation process being taken as the time at which the valve was sealed. After 7 days, the chamber was backfilled to 760~Torr with dry nitrogen by temporarily attaching a compressed nitrogen cylinder, and the RAD7 was connected to the chamber. The inlet and outlet valves were then opened, exposing the RAD7 to the emanation vessel, and data acquisition commenced. Data analysis proceeded exactly as in section~\ref{sec:rad7}, except that the emanation vessel now contained a sample as well as the gas to be analysed. This volume reduction was accounted for in analysis.

\subsection{Radon emanation results from materials}
Table~\ref{tbl:radonresults} summarises the main results from suspected radon-hot detector components and their replacements, acquired in the surface laboratory at the University of Sheffield and underground at Boulby between 2006 and 2012. Example background measurements are also presented, along with measurements of a set of ribbon cables - a known radon-emitting component from a previous detector (DRIFT-IIb) - as a consistency check. The ribbon cables were tested for radon emanation in two different vessels with separately measured backgrounds, and the results were found to be in agreement. Repeatability was verified by measuring the aluminised Mylar cathode material three times with exactly the same setup.  Suspected radon emanators present during the example run of section~\ref{sec:2.2} appear in Table~\ref{tbl:radonresults} with selected replacement materials immediately below them. The full-detector radon emanation rate measured in section~\ref{sec:rad7} is included at the bottom of the table. 

\begin{table}
\begin{center}
\resizebox{\textwidth}{!} {
\begin{tabular}{|c|c|c|c|}
\hline
Sample 													& Vessel Volume (L)	& Sample Dimensions  								& Rn emanation (atoms/min) \\ \hline
Example background										&	35 			&	N/A											&	$1.56 \pm 0.30$ \\ 
Example background 										&	3.5			&	N/A									 	 	& 	$0.63 \pm 0.08$ \\ \hline
Ribbon cables		 										&	35 			&	$2.56 \pm 0.07 $~\si{\kilogram}				 	 	& 	$24.40 \pm 0.67$ \\ 
Low-Pb ribbon cables 										&	35 			&	$\sim 2.56$~\si{\kilogram}							&	$8.40 \pm 0.60$ \\ 
FEP ribbon cables											&	35 			&	$\sim 2.56$~\si{\kilogram}							& 	$< 1.20$ \\  
Thermoplastic (TPE) ribbon cable roll (3M 3749/50, 3749/100)			&	35 			&	l: $30$~m. Wound roll V: $2.23$~\si{\litre} 			&	$< 1.20$ \\ 
Electronics Boxes 											&	35 			&	-										 	& 	$3.00 \pm 0.60$ \\ 
DRIFT-IIb Grouping boards 									&	35 			&	-										 	& 	$ < 1.20$ \\ 
DRIFT-IIb Field cage parts 									&	35 			&	-									 	 	& 	$ < 1.80$ \\ 
RG58 signal cables (standard insulation) 						&	35 			&	l: $72$~m										&	$21.60 \pm 1.80$ \\ 
RG58 signal cables (PTFE insulation) 							&	35 			&	l: $72$~m								 	 	& 	$< 1.20$\\ 
Nitrile O-ring (DRIFT vessel door seal) 							&	35 			&	ID: $6.42$~m. D: $9.5$~mm						&	$7.21 \pm 0.70$ \\ 
Teflon-encapsulated Viton 75 O-ring (valleyseal.com, half full length)	&	3.5 			&	ID: $3.2$~m. D: $10$~mm				 		& 	$< 1.20$ \\
HV cables (std. insulation. Plugs: Multicomp SPC15135)				&	35 			&	l: 25~m. D: $3.0 \pm 0.5$~mm					 	& 	$7.14 \pm 0.85$ \\ 
HV cables (Teflon insulation. Plugs as above)  					&	3.5 			&	l: $25$~m. D: $3.0 \pm 0.5$~mm			 	  	& 	$< 1.20$ \\
10 fluoroelastomer bungs (truncated cones)						&	3.5			&	D$_1$: $16$~mm. D$_2$: $12$~mm. h: $25$~mm	 	& 	$4.15 \pm 0.20$ \\
20 silicone bungs (Viadon LLC STP0 tapered plugs)				&	3.5 			&	D$_1$: $16$~mm. D$_2$: $12$~mm. h: $25$~mm 		& 	$1.52 \pm 0.09$ \\ 
Aluminized $0.9$~\si{\micro \meter} Mylar sheet					&	3.5			&	$0.59$~g										&	$< 0.4$ \\
4 two-layer PCBs (made from FR-4)								&	3.5 			&	$150 \times 100 \times 1.5$~mm					&	$< 0.4$ \\
4 Kapton flex PCBs w/plastic plugs (allflexinc.com)					&	3.5			&	Each $365 \times 65$~mm						&	$< 0.4$ \\
5 Lexan blocks (eplastics.com)									&	35			&	Each $\sim$ h: $10$~mm, w: $155$~mm, l: $155$~mm	&	$< 1.20$ \\
Epoxy-laminated Kevlar (Aeropoxy ES9209, Kevlar 49 style 120)		&	35			&	h: $1$~mm, w: $166$~mm, l: $166$~mm				&	$< 1.20$ \\
HHV cable (AWM3239 - 40kV 150$^\circ$C)						&	3.5			&	D: $5.9$~mm, l: 12.5~m							&	$< 0.4$ \\
HV putty (GC Electronics 10-8882)								&	3.5			&	227~g roll										&	$< 0.4$ \\
DRIFT-IIe empty vessel (stainless steel)							&	35			&	$1.5$ by $1.5$ by $1.5$~m cube					&	$4.31 \pm 0.35$ \\ \hline 
DRIFT-IId	- complete detector (see section \ref{sec:rad7})				&	35 			&	N/A									 		& 	$15.42 \pm 1.32$ \\ 
DRIFT-IId	(zero emanation time)								&	35 			&	N/A											& 	$0.28 \pm 0.79$ \\ \hline
\end{tabular}
}
\caption{Results of Rn emanation tests on DRIFT-IId detector material samples. Two example background measurements (see text) are presented, although many more were taken for each vessel. The radon emanation sensitivity (lowest level detectable) was $1.2$~atoms/min.}
\label{tbl:radonresults}
\end{center}
\end{table}

The nitrile o-ring door seal of the DRIFT-IId detector had long been suspected as a source of radon, and this was quantified by these measurements at $7.21 \pm 0.70$~atoms/min. Similarly, a set of 10 bungs of an unknown fluoroelastomer composition were tested and together found to emit $4.15 \pm 0.20$~atoms/min. Twenty of these bungs were used to seal unused holes in the two signal feedthrough flanges on top of the DRIFT-IId vessel. The HV distribution cables were also found to be a source of Rn in the vessel, with an emanation rate of $7.14 \pm 0.85$~atoms/min. The aluminised Mylar sample was from the same batch as used for the cathode operating on the DRIFT -IId detector from March 2010 to April 2012, and was found to emanate $0.35 \pm 0.07 $~atoms/min, below the sensitivity of the apparatus.  A polytetrafluoroethylene (PTFE) encapsulated o-ring was obtained to replace the nitrile one. It is well known that PTFE has very low gas permeability~\cite{Park2004} and it was hoped that combining this with the elasticity of the elastomer core would produce an o-ring that could provide both low radon emanation and an effective vacuum seal. This measurement shows a factor $\sim 10$ lower rate than its nitrile predecessor, although the exact improvement is uncertain since the measured level is below the $1.2$~atoms/min sensitivity of the apparatus. Silicone was chosen as the material to replace the fluoroelastomer bungs, due to its elasticity and low radon permeability~\cite{Keller2001}. A full set of $20$ replacement bungs were tested, and found to have a radon emanation rate five times lower than the old bungs ($1.52\pm0.09$ compared with $8.30\pm0.40$~atoms/min for an equivalent number of fluoroelastomer bungs).  A new set of HV distribution cables was manufactured, using a radio-pure Teflon insulating sheath. The completed cables were tested for radon emanation, and found to be lower by a factor of $\sim 20$ than the old ones.

\subsection{Comparison with radon rates in DRIFT}

It is interesting to ask how much of the total emanation calculated in section~\ref{sec:rad7} can be accounted for by the identified sources of radon. If the sum of individual contributions falls short of the total as measured, this would suggest that there are still more significant emanating materials in the detector. To address this the radon emanation rates from the relevant components in Table~\ref{tbl:radonresults} were scaled according to the estimated fraction of their surface area in contact with the vessel interior during experiment operation, and then summed. For example, the o-ring door seals were assumed to present the inner half of their surface area to the vacuum, yielding a scaling factor of $0.5$. The scaling was based upon sample surface area, rather than volume, because the range of the $^{222}$Rn recoil following a $^{266}$Ra decay is $\mathcal{O} (10$~nm$)$ in a solid, which prevents radon from the decay of more deeply-buried radium from emanating~\cite{Sakoda2011}. Applying the scaling to all the measured components present during the GPCC measurement of section~\ref{sec:2.2} and summing yielded a value of $24.01^{+4.39}_{-1.19}$~atoms/min, where the large upper uncertainty is generated by adding a contribution equal to the measurement sensitivity ($1.2$~atoms/min and $0.4$~atoms/min for the 35~L and 3.5~L vessels, respectively) for each of the components that were measured, but found to be emanating at a rate lower than the sensitivity of the apparatus. The total emanation rate is in excellent agreement with the value of $22.34 \pm 4.06$~atoms/min total measured from the GPCCs.

A final $13.9$~live-days dataset was collected in December 2012 after installing the replacement silicone bungs and low-radon HV distribution cables, which yielded a reduced $^{222}$Rn GPCC alpha rate of $35.7 \pm 1.6$~events / day.  Repeating the analysis of section~\ref{sec:2.2} gave a GPCC-derived radon emanation rate of  $12.42 \pm 2.25$~atoms/min. Once again scaling the emanation rates of the relevant components in Table~\ref{tbl:radonresults} by the surface area in contact with the vacuum, and summing the contributions, yields a total radon emanation rate of $12.00^{+5.18}_{-0.78}$~atoms/min, which is again in excellent agreement with the GPCC-implied rate.

The factor $\sim 2$ reduction in radon emanation reduces the rate of DRIFT's RPR background~\cite{Burgos2007} by the same factor which, to first order, increases the limit setting power of the experiment by a factor of two over that found in Daw et al. (2012)~\cite{Daw2012}. In fact, a technique for fiducialising events in the $z$-dimension has recently been discovered by Snowden-Ifft~\cite{Snowden2013a}, which allows extremely efficient rejection of the RPR background, and the resulting improvement in limit setting power will appear in a forthcoming paper. Efforts to suppress the RPR background using a novel design of alpha-transparent central cathode will be presented in a second forthcoming paper.

\section{DRIFT as a Radon Assay Facility}
\label{sec:facility}

The analysis of the preceding sections confirms that DRIFT data can be used to reliably measure radon levels in the detector. This opens up the possibility of using DRIFT itself as a radon assay instrument. DRIFT's sensitivity to radon (that is, the smallest concentration of added radon detectable by the experiment above background) was therefore determined according to the following procedure. An available livetime of 7 days was assumed for the purposes of this calculation.

The raw rate of GPCC events detected in the fiducial volume during the December 2012 run was multiplied by the assumed livetime to give the expected number of background GPCC events during a one-week long assay run, and the square root taken to give the Poisson noise on this count: $N_p = \sqrt{35.7 \text{ events/day} \times 7 \text{ days}} = 15.8$. This is the minimum number of counts that must be produced by an introduced sample in order for its radon emanation rate to be detectable above background. Using the previously-determined efficiency factor, this count was converted into the number of radon atoms that decayed in the vacuum vessel during the week-long run. Finally, dividing by the livetime in seconds and total detector volume in litres yields DRIFT's radon sensitivity: $S_D = 2.5$~\si{\micro \becquerel \per \litre}. Comparing this with the state-of-the-art RAD7's value of $100$~\si{\micro \becquerel \per \litre} \cite {Grodzins2009}, it can be seen that DRIFT has around a factor of $40$ superior sensitivity. For a sense of scale, the radon activity in the air of the Boulby Underground Laboratory (radon-cold compared with typical surface locations) was measured to be $3730 \pm 790$~\si{\micro \becquerel \per \litre}, much lower than at the Gran Sasso laboratory ($20000-80000$~\si{\micro \becquerel \per \litre})~\cite{LNGS} .

\section{Conclusions}
Measurements of the radon emanation rate from detector materials into the DRIFT-IId vacuum vessel have been made using a RAD7 $\alpha$~spectrometer. The results ($15.42 \pm 1.32$~atoms/min) are slightly lower than those yielded by an analysis of contained GPCC events in DRIFT data ($22.34\pm4.06$~atoms/min), which are an unambiguous tracer of radon. However, the results are in remarkably good agreement considering the techniques are completely independent of one another. Individual detector components, both existing and proposed, were also tested separately for radon emanation using a similar apparatus. A comparison of the summed rates from those that were present in the detector ($24.01^{+4.39}_{-1.19}$~atoms/min) with the overall radon emanation rate ($22.34\pm4.06$~atoms/min) suggests that all significant contributors of radon in DRIFT-IId have been identified, and in many cases these have been replaced with radon-cold alternatives. This materials substitution programme, along with other radon mitigation measures, has resulted in an overall reduction in the rate of radon emanation into the DRIFT vacuum vessel by a factor of $10$ since 2005, with a corresponding improvement in the experiment's limit setting power by a similar factor. The feasibility of screening materials to minimise the radon emanation rate in DRIFT-IId has been shown. A dataset was analysed for the radon emanation rate after substitution of materials, which was found to be $12.42 \pm 2.25$~atoms/min.  This was again found to be in excellent agreement with the sum of the remaining identified radon emanators present inside the vacuum vessel ($12.00^{+5.18}_{-0.78}$~atoms/min). Finally, the intrinsic sensitivity of a suggested radon assay facility based upon the DRIFT-II detector was estimated to be 2.5~\si{\micro \becquerel \per \litre}.  Further work on materials screening and substitution is ongoing with another factor $2$ reduction in the radon emanation rate expected.
\acknowledgments
The authors would like to acknowledge the support of the NSF under grant numbers 1103420 and 1103511, as well as the mine company CPL.  S. Sadler was supported on an STFC PhD grant.

\end{document}